\documentclass[%
 reprint,
 amsmath,amssymb,
 aps,
 prc,
 floatfix,
]{revtex4-1}

\usepackage{graphicx}
\usepackage{dcolumn}
\usepackage{bm}

\begin{document}


\title{Understanding fission fragment mass distributions\\in a shape modified random neck rupture model}

\author{Y. Sawant}
\email{ysawant@barc.gov.in}
\author{S. V. Suryanarayana}
\author{ B. K. Nayak}  
\author{ A. Saxena} 
\author{R. K. Choudhury}

\affiliation{Nuclear Physics Division, Bhabha Atomic Research Centre, Mumbai - 400085, INDIA}

\date{\today}

\begin{abstract}
The variances of the fission fragment mass distributions for symmetric case, over a 
wide range of the fissility of the compound nucleus have been investigated within the 
frame work of Random Neck Rupture Model (RNRM) proposed by Brosa et al.. The shape 
of the fissioning nucleus is generated excluding c$_{rel}$ (i.e.c$_{rel}$=1) in the RNRM model,  which  results  in more continuous shape of fissioning nucleus at boundaries connecting heads to neck in scission shape. This shape modified RNRM  model  has been used to analyse experimental data of mass variances for 
symmetric mass distributions of 27 systems in a wide region of fissility 0.7-0.95. The neck radius is  an important parameter of the RNRM model and this has been varied to fit the experimental mass  variances data. The systematics of resulting neck radii are studied of as a function of fissility and nuclear potential through $\gamma_0$, the surface energy coefficient. Further average total kinetic energies of fission fragments $<$TKE$>$ are studied of as a function of fissility and $\gamma_0$ parameter. It is found that the neck radii that fit the experimentally observed variances of the mass distributions fall into two groups and these groups are related to two groups 
of experimentally observed  $<$TKE$>$s. Empirical formulae have been obtained for the neck 
radii for these two groups of fissioning systems.  Use of the empirical formulae for  
neck radii predict the mass variances reasonably well for five test systems and it is shown that these empirical neck radii are better estimates than starting with 
the Rayleigh criterion.

\end{abstract}

\pacs{Valid PACS appear here}
\maketitle


\section{Introduction}

Fission fragment mass distribution is an important observable of the fission process 
that results from shape evolution dynamics of fissioning nucleus. This is in general 
true for fission induced by any projectile such as neutron, proton, ${\alpha}$ or heavy ions. 
It is known that the mass distribution from fission of a fully equilibrated compound 
system is independent of projectile or entrance channel and depends on the excitation 
energy and spin of the fissioning nucleus. Studies on the mass distributions provide 
valuable information about the potential energy landscape of the fissioning nucleus 
and the complex fission mechanism exhibiting several effects such as symmetric or 
asymmetric mass distributions, peak to valley ratios, fission fragment angular anisotropy, 
total fragment kinetic energies, emissions such as neutrons, ${\gamma}$ and ${\alpha}$  during or after 
scission process known as pre-scission or post-scission multiplicities \cite{Vande,Wagge}. 
These multiplicities and other fission observables are important for understanding the 
fissioning system shape evolution by means of fission models such as Brosa model or Langevin dynamics.\\ 

	A large amount of experimental data on the mass distribution in nuclear fission has been generated over
 the years. Early studies on low-energy fission of actinides revealed the importance of the nuclear 
shell effects in fission. The main interest in the medium-energy heavy-ion-induced fission is to 
understand the effects of entrance channel parameters namely, projectile energy, angular momentum 
and entrance channel mass asymmetry, on the fission process. An analysis of the data on the variance 
of the mass distribution over a wide range of the fissility of the compound nucleus was reported in 
many publications \cite{RusaZPhys,YsawPRC,RkcPRC,ShawPRC,ShenPRC,BerrNature,ChubPAN,ZhdaSJNPA}. The 
analysis revealed that the variance of the mass distribution shows fissility (${ \chi}$) dependence when studied 
as a function of T$_{CN}$ which represents temperature at the saddle point.  However, fissility dependence 
vanishes when studied as a function of fragment temperature T$_F$  \cite{YsawPRC}, that corresponds to 
scission point temperature T$_{sc}$. Thus, the variance of the mass distribution provides important 
information about the fission process and can be used to  test various models of fission such as 
the saddle point model \cite{NixNPA} and the scission point model \cite{TsaNPA,WilkPRC,BrsPRL,BrsRep}. 
These models, although qualitatively explain the gross features of the mass distributions, fail to 
quantitatively explain the mass distributions.\\
 
	Brosa et al. \cite{BrsRep} proposed the random neck rupture model (RNRM) for the calculation of 
post-fission observables such as mass distribution, kinetic energy distribution and neutron 
multiplicity. According to this model, the pre-scission shape of the fissioning nucleus dictates 
the post-fission observables. This model has been successful in explaining the width of the mass 
distribution in low- as well as medium-energy fission \cite{BrsPRC}. In the present work, experimentally 
determined variances of the symmetric mass distributions have been compared with the results from using 
the modified shape in RNRM  for 27 systems, data taken from 
\cite{YsawPRC,RkcPRC,ShawPRC,ShenPRC,BerrNature,ChubPAN,ZhdaSJNPA,RamEPJW,PrasPRC,ItkiPRC,ProkNPA,NishPRC}.

\section{Shape evolution in Brosa model}

According to  Brosa et al. RNRM model \cite{BrsRep}, the compound nucleus undergoes a shape change 
from a near spheroidal shape at the saddle point to an elongated deformed shape, called a prescission 
shape, which is the last stage before the neck ruptures. This shape is normally described by a long flat 
neck  connecting two spherical heads. In this model, during the motion of the fissioning nucleus towards 
scission, a dent is developed in the neck region and is deepened by the capillary force finally leading to 
fission. The curvature of the fissioning nucleus changes from positive to negative in the motion towards 
scission. During this transition when the neck becomes flat, there can be a large shift in the position 
of the dent without sizeable physical mass motion, which  finally leads to large mass fluctuations in the 
fission process. In the RNRM model \cite{BrsRep} the pre-scission shape of fissioning nucleus is described 
by the following set of equations, suitable for symmetric fission.\\

\begin{equation}
\rho(\zeta) = \left\{ \begin{array}{cc}
 \sqrt{r_1^2-\zeta^2}  & ~~ -r_1 \le \zeta \le \zeta_1  \vspace{0.2cm}\\ 
 r+a^2c [cosh(\frac{\zeta-z}{a})-1] & \zeta_1 \le \zeta \le \zeta_2 ~~ \vspace{0.3cm}\\
 \sqrt{r_1^2-(2l-2r_1-\zeta)^2} & ~~~~ \zeta_2 \le \zeta \le 2l-r_1  
\end{array} \right.
\end{equation}

Equation (1) represents a shape that is made up of two spheres connected by a neck with minimal 
curvature 'c'. For symmetric case shape made up of two equal spheres can be assumed.\\

\begin{figure}
\includegraphics[width=2.7in]{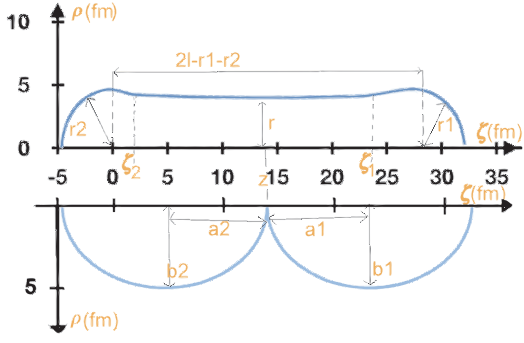}
\caption{ The shape shown here is a prescission shape with flat-neck representation, upper part, and the embedded spheroids, lower part. Lengths are measured in fermi  and should be realistic with an accuracy of 10$\%$ }
\end{figure}

	The shape in RNRM model of Eq.(1)  is shown schematically in Figure 1. In this work, we consider 
symmetric fission cases and there  are six parameters (r$_1$, $\zeta_1$, r, a, c, l) for this 
shape. Here, 'r$_1$' is taken as radius of the spherical heads at both ends of the pre-scission 
shape (for symmetric case, r$_1$= r$_2$ ), 'r' is the minimal neck radius ($r_{neck}$) and   ${\zeta}_1$ is 
the transitional point  where the function describing the shape changes. Also, the position of  
${\zeta}_2$  is calculated using   $\zeta_1$, hence eliminated from parameters list. 'c' is the 
curvature of the neck, where the neck is thinnest, i.e. at the geometrical centre of the shape 
in the case of symmetric pre-scission shape. The parameter 'a' is a measure of the extension of 
the neck and '2l' is the total elongation of the pre-scission shape. By imposing the conditions 
of continuity of the shape and volume conservation, a set of nonlinear equations were solved to 
determine r$_1$, r, $\zeta_1$, a and 'c'. Further the parameter 'c' can be  correlated 
to r$_1$ , r and  l  using the following equation:

\begin{equation}
c=2c_{rel}\left(\frac{r_1-r}{l-r_1}\right) 
\end{equation}

The value of c$_{rel}$ was taken so as to z and dz/dr become continuous at the transitional 
points z1, z2 as mentioned in \cite{BrsRep}, which gives continuous shape as well as 
continuously differentiable shape. \\

Further we modified shape of  Brosa's RNRM model by excluding c$_{rel}$ parameter in curvature formula (i.e. taking c$_{rel}$ =1) resulting in more continuous shape at boundaries connecting heads to neck in scission shape. Remaining variable 'l' was varied to reproduce the experimental 
average total kinetic energy  $<$TKE$>$, taken from Viola systematics \cite{VioPRC}.  For a given 
value of  l, the pre-scission shape was determined and the probability of neck rupture at 
different positions of the neck (z$_r$) was calculated using the following Eq. 3.

\begin{equation}
W(A)=exp\left(-2\pi\gamma_0\frac{[\rho^2(z_r)-\rho^2(z)]}{T_{sc}}\right) 
\end{equation}
    where,
\begin{equation}
\gamma_0=0.9517 \left[1-1.7828 \left(\frac{N_{CN}-Z_{CN}}{A_{CN}}\right)^2\right]
\end{equation}

In Eq. 4, N$_{CN}$ , Z$_{CN}$ and A$_{CN}$ are the neutron number,  atomic number and mass number of the 
fissioning nucleus and T$_{sc}$ is the temperature of the fissioning nucleus at the scission point. 
The elongated deformed nucleus at the scission point splits into two deformed fragments and the 
deformation energy of these nascent fragments gets added to the excitation energy of the fission 
fragments. This excitation energy is released by evaporation of neutrons from the fragments, 
which is normally the measure of the temperature of the fragments. The temperature  T$_{sc}$  at 
the scission point was calculated using formula T$_{sc}$ = ${\sqrt{E^*_{sc} /a}}$, where E$^*_{sc}$    
is the excitation energy of the fissioning nucleus at the scission point, '$a$'  is level density 
parameter at scission point. The scission point excitation energy (E$^*_{sc}$) in terms of 
fragment's energy (E$_{Frag}$) after scission and deformation energy (E$_{def}$) at scission point 
is given by, E$^*_{sc}$ = ( E$_{Frag}$  -  E$_{def}$ ) A$_{CN}$ /A$_{Frag}$. 
The excitation energy of fragments E$_{Frag}$  after the scission is 
calculated using formula E$_{Frag}$  = $a^2{T_{F}}$, where T$_{F}$    
is the temperature of the fragment, and '$a$'  is level density 
parameter for fragment. The temperature of fragment (T$_F$) after scission is 
calculated  using method described in Ref. \cite{SaxePRC}.\\

The random neck rupture produces different mass fragments by chopping the neck at 
different positions. So one can say that the prescission shape is related to the 
width of mass distribution. More precisely, the variance of the mass distribution 
strongly depends on the neck of scission point shape. Also, if the temperature at 
the scission point is high, the fluctuation in the rupture position will be larger 
and it will give rise to a broader mass distribution.  The scission configuration, 
which includes the scission excitation energy and scission point deformation energy, 
plays an important role in deciding the width of the mass distribution. The rupture 
position (z$_r$ ) was translated into the fragment mass using the following relation:

\begin{equation}
A(z_r)=\left(\frac{3A_{CN}}{4r^3_{CN}}\right){\displaystyle \int_{-r_1}^{z_r} \rho^2(\zeta)d\zeta} 
\end{equation}

\begin{figure}
\includegraphics[width=2.7in]{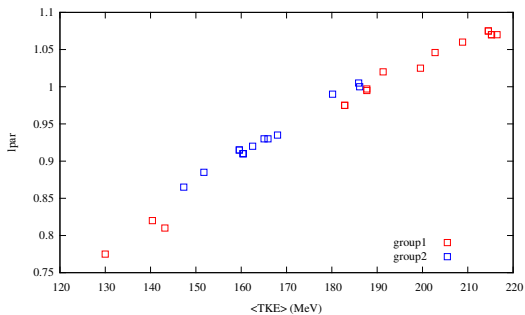}
\caption{(Color online) $lpar$ representing scission semilength ($l$)  versus average total kinetic energy of fragments $<$TKE$>$ from Viola systematics \cite{VioPRC} from Brosa model calculations.}
\end{figure}

As mentioned earlier, we fix c$_{rel}$ =1 and adjust  elongation length and neck radius $r_{neck}$ 
to fit experimental data. The total length of the fissioning shape is maximum about 40fm and 
we perform shape calculations in step size of 0.02fm in order to smoothly join the heads and 
neck in the overlapping regions giving a smooth and continuous shape. The average total 
fragments kinetic energy strongly depends on total elongation length '2l' and  here we 
introduce a parameter $lpar$=l/(r$_0$*15) fm (with fixed r$_0$=1.2249 fm). In the present RNRM 
calculations, the elongation '$lpar$' of the pre-scission shape was varied to reproduce 
the experimental average total kinetic energy $<$TKE$>$ for given compound nucleus and as 
well as the shape at scission point.  Figure 2 shows typical dependence of $<$TKE$>$ versus 
these adjusted elongation $lpar$ values.  It can be seen that $<$TKE$>$ shows a linear dependence 
on elongation (in figure $lpar$ is shown) for all fissioning systems irrespective of entrance channel. \\

\begin{figure}
\includegraphics[width=2.7in]{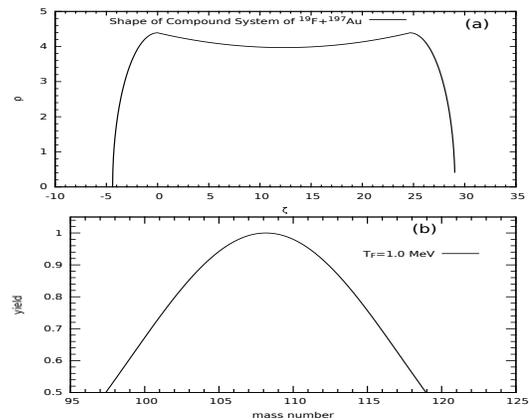}
\caption{(Color online) (a) Shape of fissioning compound system formed 
in $^{19}$F+$^{197}$Au using Brosa Model. (b) Fission fragments mass distribution from  
corresponding shape  shown in Fig. 3(a).}
\end{figure}

Using this information of scission length for the fissioning nucleus from the experimental $<$TKE$>$ 
data, the scission point shape of the fissioning nucleus of Brosa's model (as  in Figure 1) can be 
configured for a fixed neck radius $r_{neck}$  and the shape is sensitive to $r_{neck}$. Further this 
shape ruptures at different points on flat neck of scissioning nucleus depending on the scission 
point temperature to produce distribution of fragment masses. The flatness and neck radius 
influence the width of distribution of fragment masses.  Eq. (3)  indicates that more the 
scission point temperature,  more will be  variance of the mass distribution for given 
compound nucleus.  This shape analyses of the fissioning nuclei were carried out for the 
27 systems with real time monitoring of shape configuration using graphic user interface 
(GUI), while the parameters $lpar$ and $r_{neck}$ were being continuously adjusted for $<$TKE$>$ and 
mass variances data. In the program, the head radius gets automatically fixed from volume 
conservation. \\

As an example of present work, we show in Fig. 3(a)  the shape of fissioning compound nucleus formed by $^{19}$F+$^{197}$Au system. This shape  is attained after adjusting the $lpar$ to fit the $<$TKE$>$ 
of Viola systematics \cite{VioPRC} and also after adjusting the $r_{neck}$ to fit the width of mass distributions 
data for this system at T$_F$=1.0 MeV. Once the shape is determined at one temperature, the mass 
distributions for different temperatures T$_F$ corresponding to respective T$_{sc}$ were calculated 
using Eq. 3. The resulting mass distribution versus mass number is shown in Fig. 3(b) 
T$_F$=1.0 MeV and the intercepts on x-axis give full width at half maximum (maximum is 1.0) 
for the mass distribution and using this the variance can be calculated.\\

\begin{table}
\caption{$r_{neck}$, ${\gamma}_0$, ${\chi}$ , $lpar$ for different systems in two groups}
\begin{tabular}{|l|c|c|c|c|c|}
\hline
System-group1		&	${\gamma}_0$ &	${\chi}$  &	$lpar$	&	$r_{neck}$ &	Ref.	\\
\hline
$^{16}$O+$^{209}$Bi	&	0.88974	&	0.77376	&	0.935	&	3.9	&	\cite{RkcPRC} \\
$^{12}$C+$^{232}$Th	&	0.87465	&	0.80778	&	0.99	&	4.008	&	\cite{RkcPRC}	\\
$^{16}$O+$^{232}$Th	&	0.87711	&	0.82588	&	1.005	&	4.0205	&	\cite{RkcPRC}	\\
$^{16}$O+$^{204}$Pb	&	0.89562	&	0.76897	&	0.93	&	3.97	&	\cite{ChubPAN} \\
$^{16}$O+$^{208}$Pb	&	0.88624	&	0.76323	&	0.93	&	4.0086	&	\cite{ChubPAN} \\
$^{30}$Si+$^{186}$W	&	0.89352	&	0.75055	&	0.91	&	3.97	&	\cite{BerrNature} \\
$^{19}$F+$^{197}$Au	&	0.89352	&	0.75055	&	0.91	&	3.97	&	\cite{BerrNature} \\
$^{12}$C+$^{204}$Pb	&	0.89352	&	0.75055	&	0.91	&	3.97	&	\cite{BerrNature} \\
$^{4}$He+$^{209}$Bi	&	0.88256	&	0.71893	&	0.885	&	3.9924	&	\cite{ZhdaSJNPA}	\\
$^{4}$He+$^{201}$Tl	&	0.8903	&	0.70605	&	0.865	&	3.9925	&	\cite{ZhdaSJNPA}	\\
$^{12}$C+$^{209}$Bi	&	0.88748	&	0.75544	&	0.92	&	4.0082	&	\cite{YsawPRC} 	\\
$^{12}$C+$^{235}$U	&	0.87937	&	0.82709	&	1	&	4.0315	&	\cite{YsawPRC}	\\
$^{12}$C+$^{208}$Pb	&	0.88384	&	0.74497	&	0.915	&	3.965	&	\cite{YsawPRC}	\\
$^{11}$B+$^{209}$Bi	&	0.88384	&	0.74497	&	0.915	&	3.965	&	\cite{YsawPRC}	\\
$^{48}$Ca+$^{208}$Pb	&	0.8983	&	0.87357	&	1.025	&	4.02	&	\cite{ProkNPA}	\\
\hline
System-group2		&	${\gamma}_0$ &	${\chi}$  &	$lpar$	&	$r_{neck}$ &	Ref.	\\
\hline
$^{238}$U+$^{16}$O	&	0.87502	&	0.84163	&	1.02	&	4.12	&	\cite{ShenPRC}	\\
$^{238}$U+$^{26}$Mg	&	0.87537	&	0.87548	&	1.046	&	4.127	&	\cite{ShenPRC} \\
$^{32}$S+$^{208}$Pb	&	0.89468	&	0.83665	&	0.997	&	4.0413	&	\cite{ShawPRC} \\
$^{26}$Mg+$^{248}$Cm 	&	0.87568	&	0.90933	&	1.075	&	4.1	&	\cite{ItkiPRC}	\\
$^{36}$S+$^{238}$U	&	0.87568	&	0.90933	&	1.075	&	4.1	&	\cite{ItkiPRC}	\\
$^{22}$Ne+$^{249}$Cf	&	0.88182	&	0.913	&	1.07	&	4.0925	&	\cite{ItkiPRC}	\\
$^{58}$Fe+$^{208}$Pb	&	0.89176	&	0.91979	&	1.07	&	4.095	&	\cite{ItkiPRC}	\\
$^{40}$Ca+$^{192}$Os	&	0.90127	&	0.82446	&	0.975	&	4.0442	&	\cite{PrasPRC}	\\
$^{40}$Ca+$^{142}$Nd	&	0.92692	&	0.70964	&	0.81	&	3.95	&	\cite{PrasPRC}	\\
$^{13}$C+$^{176}$Yb	&	0.88668	&	0.64472	&	0.775	&	4.0354	&	\cite{RamEPJW}	\\
$^{13}$C+$^{182}$W	&	0.89705	&	0.68439	&	0.82	&	4.0262	&	\cite{RamEPJW}	\\
$^{30}$Si+$^{238}$U	&	0.87763	&	0.89359	&	1.06	&	4.15	&	\cite{NishPRC}	\\
\hline
\end{tabular}
\end{table}

\begin{figure}[b]
\includegraphics[width =2.7in]{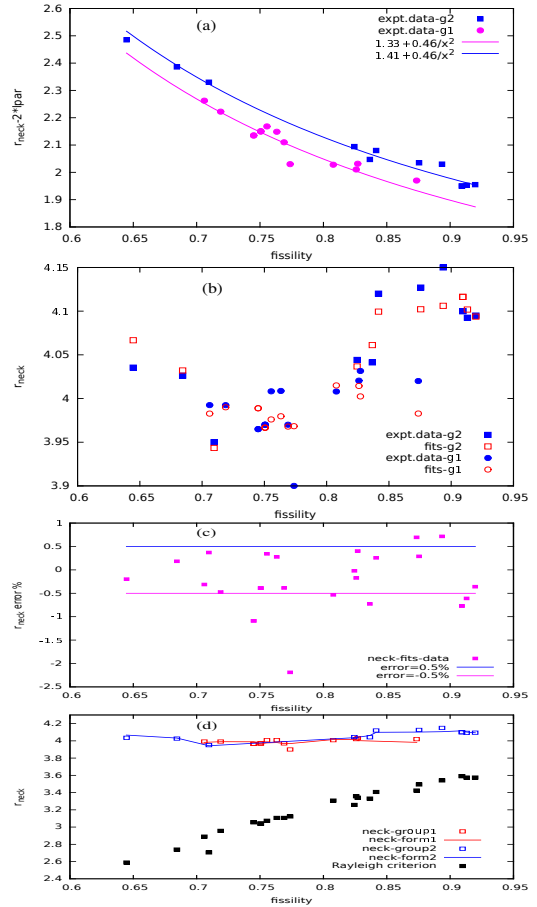}
\caption{\label{fig:epsart} (Color online) (a)  variable ($r_{neck}$ - ${2*lpar}$) vs fissility (${\chi}$) with empirical 
fits for two groups in blue and pink colour. \\ 
(b) $r_{neck}$  vs (${\chi}$)  with blue colour for neck radii from Table 1 column 5 and  pink colour  
for radii obtained using fitted parameters of Fig. 4(a).\\
(c) $r_{neck}$  error in ${\%}$ vs (${\chi}$). Straight lines are limits 
for  +0.5${\%}$  error  (blue) and -0.5${\%}$  error (pink) in radii for fitted Eqns. \\
(d) Variation in $r_{neck}$ from empirical values (lines) and data 
from Table 1. (symbols), as a function of (${\chi}$) as compared to $r_{neck}$ 
values from Rayleigh criterion (filled black square) \cite{BrsRep}. }
\end{figure}

Table 1 shows the list of systems of symmetric fission cases  studied, the ${\gamma}_0$  and (${\chi}$) parameters of the compound systems,  the $lpar$ values that fit $<$TKE$>$ data, 
$r_{neck}$ that fit the experimental variances data and references. The experimental mass 
variances are very well reproduced by varying neck radii of scission shape within 
modified shape of Brosa's RNRM model and the best fit values are shown in column 5 of 
the Table. 1. The detailed study of effect of various parameters such as fissility (${\chi}$), 
average total kinetic energy of fission fragments $<$TKE$>$ through $lpar$, nuclear 
potential through, ${\gamma}_0$ (surface energy coefficient) parameter and entrance channel, 
on the neck radius ($r_{neck}$,) is discussed below.

\section{Systematics of various parameters of Brosa model}

\begin{figure}
\includegraphics[width=2.7in]{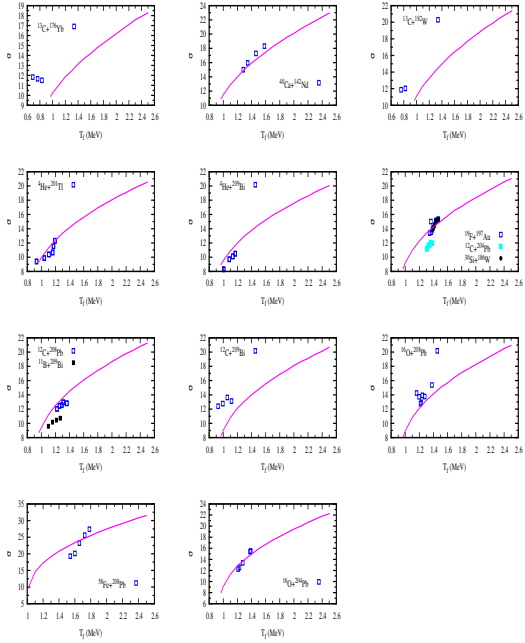}
\caption{(Color online) (a) mass variances versus T$_F$ using shape modified RNRM 
using two empirical formulae for neck size, as discussed in Fig. 4. }
\label{fig5a}
\end{figure}
\begin{figure}
\setcounter{figure}{4}
\includegraphics[width=2.7in]{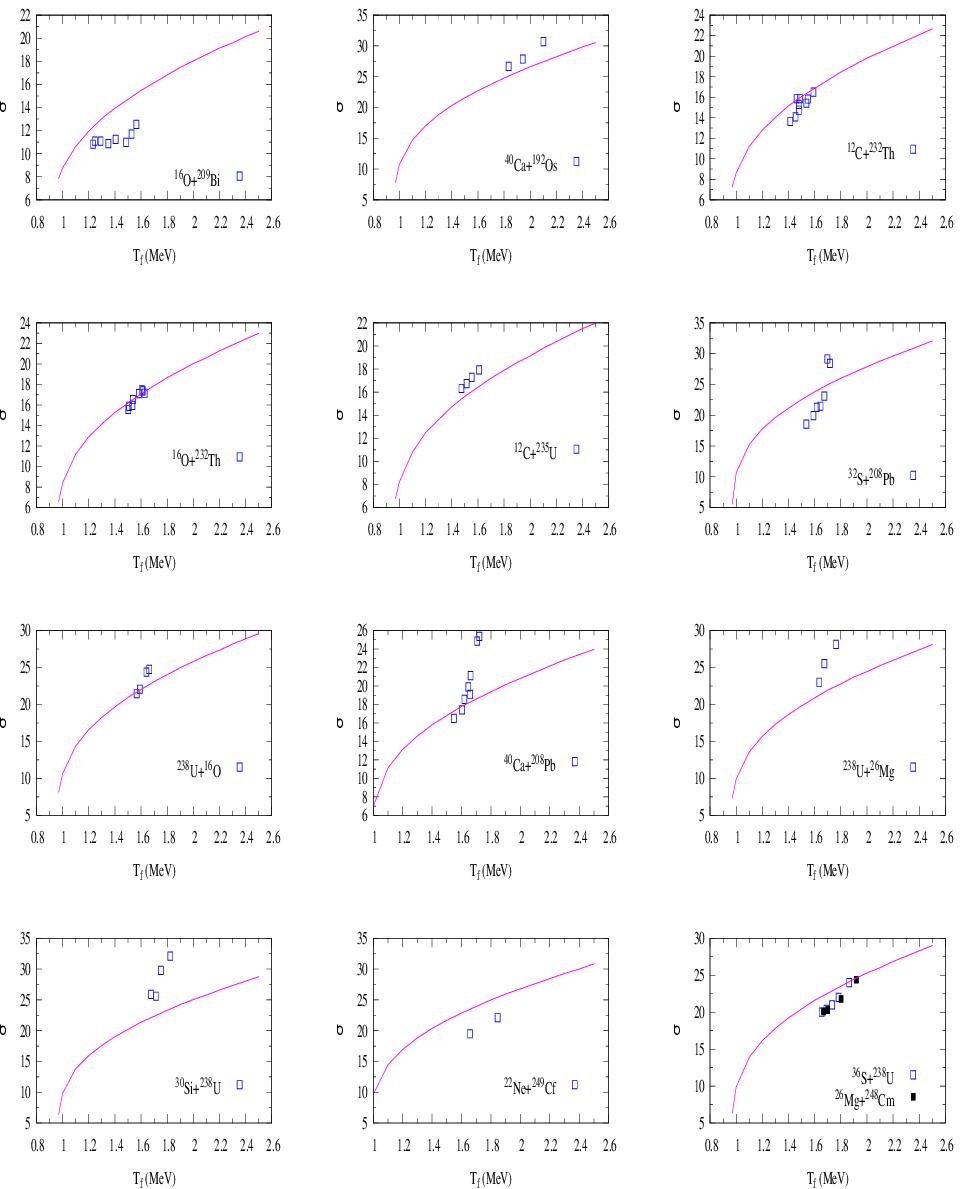}
\caption{(Color online) (b) mass variances versus T$_F$ obtained similar to Fig. 5(a).\\
\\}
\label{fig5b}
\setcounter{figure}{4}
\includegraphics[width = 2.7in]{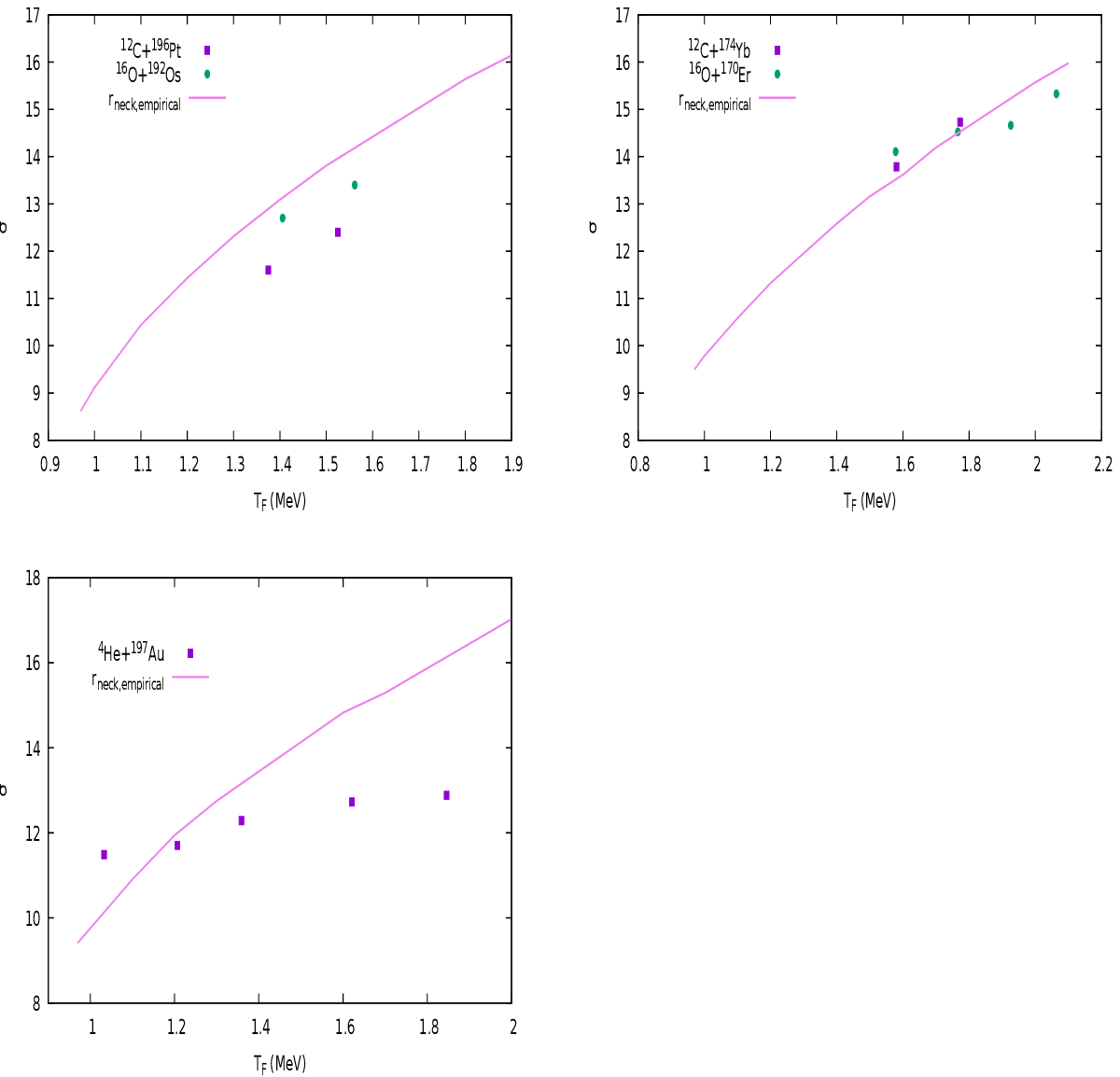}
\caption{(Color online) (c) mass variances versus T$_F$ using shape modified RNRM 
using two empirical formulae for neck size, as discussed in Fig. 4(a). and selection of value of 'k' parameter using grouping shown in Fig. 6(b) for five test systems from Refs.\cite{CuniIAEASymPhyChem,PlasPR}, 
below to show predictive power of two-fit formula. }
\end{figure}

As mentioned before, to reproduce experimental mass variances \cite{YsawPRC,RkcPRC,ShawPRC,ShenPRC,BerrNature,ChubPAN,ZhdaSJNPA,RamEPJW,PrasPRC,ItkiPRC,ProkNPA,NishPRC}, the neck radii ($r_{neck}$) have been adjusted.  The resulting $r_{neck}$ values have been analysed as a function of fissility, ${\gamma}_0$ and other parameters. The fitted r$_{neck}$ values do not show any systematic behaviour. 
However, it is observed that the quantity ($r_{neck}$ -$2*lpar$) versus fissility (${\chi}$) shows very good systematic trend and this quantity falls into two groups as a function of (${\chi}$), as shown in Fig. 4(a). The two groups in Fig. 4(a) with blue and pink symbols, can be fitted with equation of the form (k+m/${\chi}^{2}$)  with two values of constant k (=1.33,1.41) with same m=0.46 value as shown by blue and pink coloured lines. \\

\begin{figure}
\includegraphics[width=2.7in]{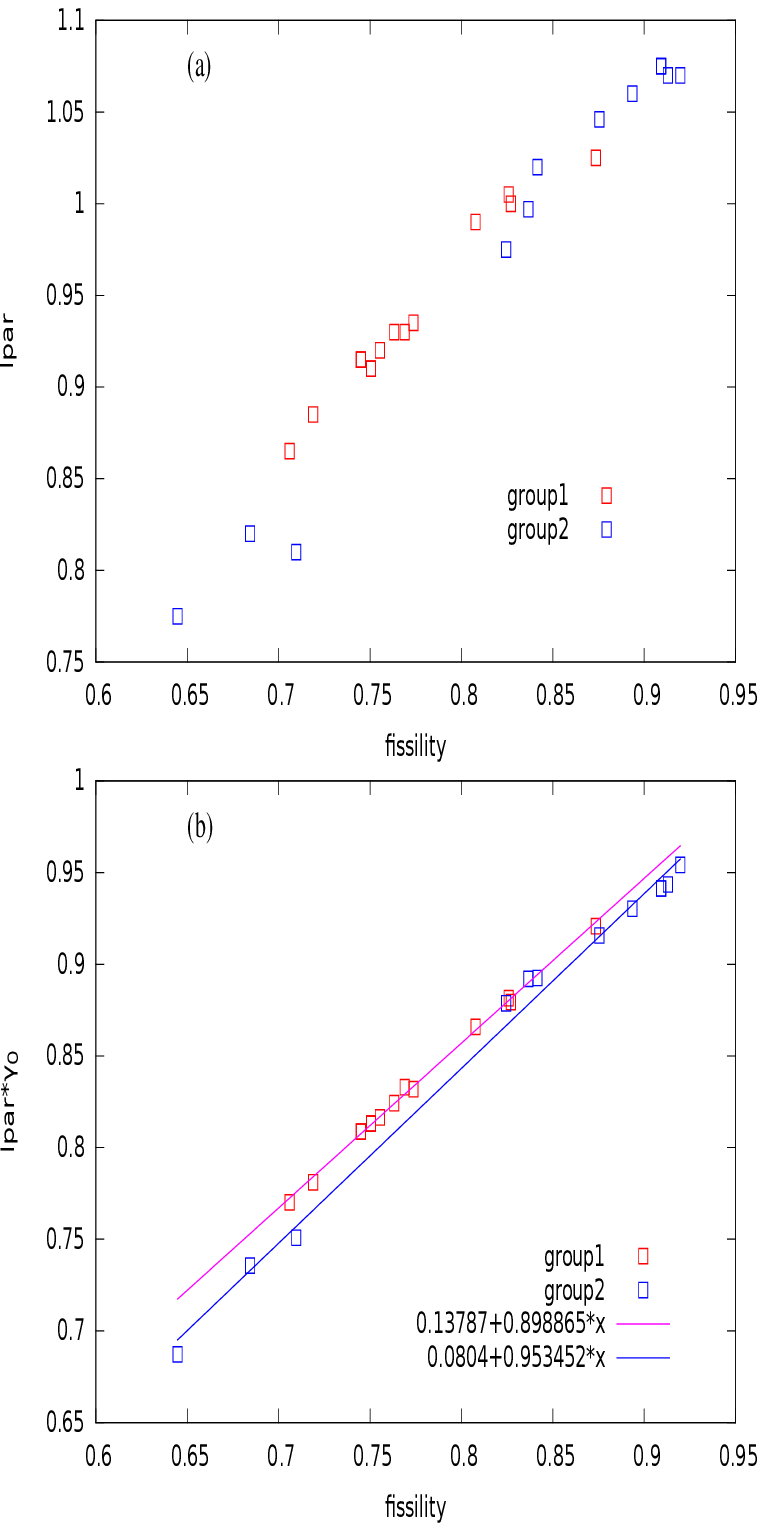}
\caption{(Color online) (a) $lpar$ vs fissility and (b) $lpar*{\gamma}_0$  vs fissility.}
\end{figure}

\begin{figure}
\includegraphics[width=2.7in]{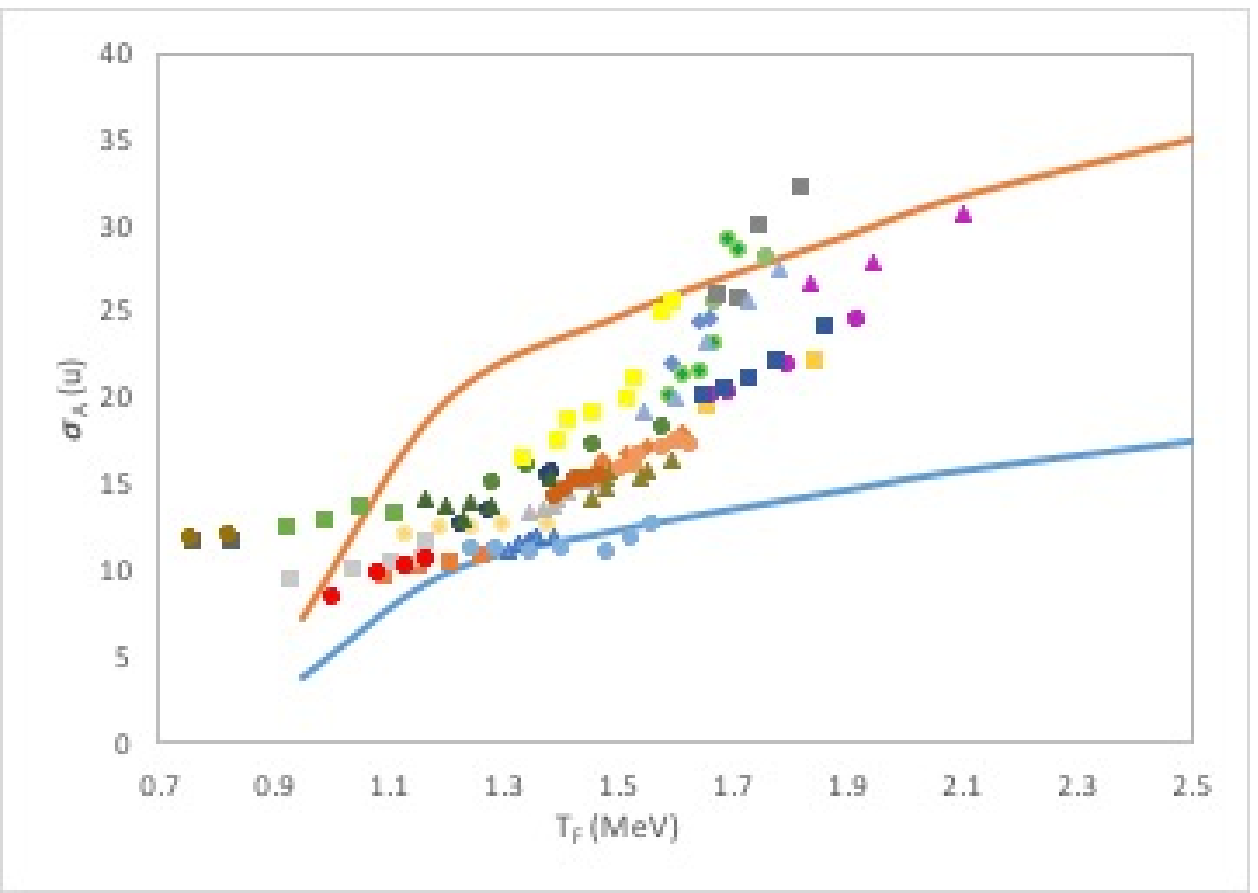}
\caption{ Symbols shows Exp. data \cite{YsawPRC,RkcPRC,ShawPRC,ShenPRC,BerrNature,ChubPAN,ZhdaSJNPA,RamEPJW,PrasPRC,ItkiPRC,ProkNPA,NishPRC} whereas lines for shape modified Brosa model calculations set up. The experimental uncertainties are within the symbol sizes.}
\end{figure}

	The empirical formulae in Fig. 4(a) give ($r_{neck}-2*lpar$) values and from these 
the $r_{neck}$ values can be obtained by adding $2*lpar$. These empirically determined neck 
radii are shown in Fig. 4(b) by pink symbols along with $r_{neck}$ values of Table. 1 in 
blue symbols. It can be seen from Fig. 4(b), the empirical  $r_{neck}$ values  are very 
close to $r_{neck}$ values of Table. 1. The percentage variation of  empirical $r_{neck}$ 
from $r_{neck}$ values of Table. 1 are shown in Fig. 4(c).  The variation is within ${\pm}0.5{\%}$, which 
can introduce error upto 2 mass units in standard deviation of fission fragment mass 
distribution. The $r_{neck}$ values from Rayleigh criterion, suggested in \cite{BrsRep} 
are compared with empirical $r_{neck}$ values as shown in Fig. 4(d). It can be seen that 
empirical $r_{neck}$ values obtained from any of two fits formulae are better estimates 
than starting with the Rayleigh criterion.\\

	Using the systematics study and the empirical formulae for neck radii, we recalculated 
mass variances values for all the 27 systems for all temperatures using RNRM model. 
Figures 5(a,b) show the experimental fission fragment mass variances data as a 
function of temperature of fragment T$_F$ along with lines predicted  using 
empirical $r_{neck}$ (from equations in figure 4(a)). All systems show
 reasonable agreement with predictions using shape modified Brosa model. 
However, the two systems $^{13}$C + $^{182}$W ${\rightarrow}$ $^{195}$Hg and $^{13}$C + $^{176}$Yb ${\rightarrow}$ $^{189}$Os 
show large deviations, may be due to shell effects as mentioned in \cite{RamEPJW} 
resulting in asymmetric fission. So, these data of the two systems cannot be analysed 
with this shape model which is mainly suitable for symmetric fission.\\

As mentioned before, the experimental $<$TKE$>$ values of Viola systematics \cite{VioPRC} have been reproduced by 
adjusting $lpar$ value in Brosa model and these $lpar$ versus fissility are shown in Fig. 6(a). As 
seen in figure, $lpar$ values exhibit  two groups, whereas no systematic behaviour was found with 
respect to either fissility or $\gamma_0$. However, when ($lpar*{\gamma}_0$) is plotted versus fissility, these 
two groups clearly separate from each other, as shown in Fig. 6(b). This effect is also 
observed in $r_{neck}$  as discussed in Fig. 4(a).  This indicates nuclear potential plays 
important role through ${\gamma}_0$ affecting mass variances  of fissioning systems. \\

To verify the validity of empirical systematic formulae, studied in present work, 
we have applied those formulae to experimental  mass variance data of five additional systems from 
Refs.\cite{CuniIAEASymPhyChem,PlasPR}.
The comparison of experimental mass distributions with theoretically predicted values of mass distributions 
are shown in Fig. 5(c) by lines. 
It can be seen from Fig. 5(c), the experimentally observed width of fission fragments mass distribution is 
very well reproduced 
by using $r_{neck}$ obtained by two-fits formulae. The value of coefficient 'k' in the two-fits formulae is 
chosen using grouping 
shown in ($lpar*{\gamma_0}$) vs fissility, plot in Fig. 6(a).  
Thus, the predictive power of empirical systematic formulae in present study, can be clearly seen from Fig. 5(c).\\ 

 Figure 7 shows mass variances for all systems studied collectively, as a function of fragment temperature (T$_{F}$). 
The calculations were carried out using shape modified RNRM Brosa Model and are shown by two theoretical lines.  
In the calculation, for each T$_{F}$, the $r_{neck}$ values are varied from mean 
$< r_{neck} >$ values by ${\pm} 2.9 {\%}$, to obtain range of mass variances values shown by 
two theoretical lines in Fig. 7. 
The mean $< r_{neck} >$ value is taken that of $^{12}$C+$^{235}$U system and the mass distributions calculated for same system  by varying 
$< r_{neck} >$ values by ${\pm}2.9\%$, as shown by lines in Fig. 7. The change in $r_{neck}$ values 
produces change in flatness of neck and thus controls mass distributions. In Fig. 7,
symbols shows experimental data \cite{YsawPRC,RkcPRC,ShawPRC,ShenPRC,BerrNature,ChubPAN,ZhdaSJNPA,RamEPJW,PrasPRC,ItkiPRC,ProkNPA,NishPRC}, 
whereas lines are results of shape modified RNRM calculations. 
The experimental uncertainties are within the symbol sizes.  

\section{Conclusion}

The variances of the fission fragment mass distributions for symmetric fission over a wide range of the 
fissility of the compound nucleus have been studied by modifying shape of the Random Neck Rupture Model 
(RNRM) of Brosa et al., by excluding the parameter c$_{rel}$ in curvature formula (i.e. taking c$_{rel}$ =1).   This shape modified RNRM  
model has been used to analyse experimental data of mass variances for symmetric mass distributions 
of 27 systems covering a  wide region of fissility 0.7-0.95. The average total kinetic energies of 
fission fragments $<$TKE$>$ have been fitted by adjusting the elongation  ($lpar$ values) of the fissioning 
nucleus shape.  The neck radius ($r_{neck}$)of the RNRM model has been varied to fit the experimental mass 
variances data. The systematics of the fitted  $lpar$ and  $r_{neck}$ values are studied of as a function of 
fissility and nuclear potential through, ${\gamma_0}$ parameter. 
It is observed that the $r_{neck}$ values fall  into two groups and these groups are related to two groups of experimentally observed $<$TKE$>$s. Empirical 
formulae have been obtained for the $r_{neck}$  of these two groups of the fissioning systems. 
The five more test systems have been used for validating systematics approach as shown in Fig. 4(a) and Fig. 6(b) of present work.
 The grouping systematic shown in Fig. 6(b) helps in selecting one equation out of two equations of two-fits empirical formulae for $r_{neck}$.
which produces mass variances which agrees reasonably well with experimental mass variances of five test systems. 
Overall study shows  that these empirical neck radii are better estimates than starting with the Rayleigh criterion.

\bibliography{publiYS}

\end{document}